\documentclass[12pt]{article}

\usepackage{amsfonts,amsmath,amssymb,bm,graphicx,hyperref,a4wide,cite,mathabx,subfigure}

\begin{document}

\title{\textbf{Magnetic helicity in plasma of chiral fermions electroweakly interacting
with inhomogeneous matter}}

\author{Maxim Dvornikov\thanks{maxdvo@izmiran.ru} 
\\
\small{\ Pushkov Institute of Terrestrial Magnetism, Ionosphere} \\
\small{and Radiowave Propagation (IZMIRAN),} \\
\small{108840 Troitsk, Moscow, Russia; and} \\
\small{\ Physics Faculty, National Research Tomsk State University,} \\
\small{36 Lenin Avenue, 634050 Tomsk, Russia}}

\date{}

\maketitle

\begin{abstract}
We study chiral fermions electroweakly interacting with a background
matter having the nonuniform density and the velocity arbitrarily
depending on coordinates. The dynamics of this system is described
approximately by finding the Berry phase. The effective action and
the kinetic equations for right and left particles are derived. In
the case of a rotating matter, we obtain the correction to the anomalous
electric current and to the Adler anomaly. Then we study some astrophysical
applications. Assuming that the chiral imbalance in a rotating neutron
star vanishes, we obtain the rate of the magnetic helicity change
owing to the interaction of chiral electrons with background neutrons.
The characteristic time of the helicity change turns out to coincide
with the period of the magnetic cycle of some pulsars.
\end{abstract}

\section{Introduction}

Studies of the chiral phenomena, apart from the purely theoretical
interest, find numerous applications in cosmology, astroparticle physics,
accelerator and solid state physics. There is a possibility to explain
the asymmetry in heavy ions collisions basing on the chiral magnetic
(CME) and the chiral vortical (CVE) effects~\cite{Kha16,ZhaWan19}.
The CME and the chiral separation effect (CSE) are supposed to be
observed in Weyl and Dirac semimetals~\cite{Arm18} and magnetized
atomic gases~\cite{Hua16}. The generation of cosmological magnetic
fields and the lepton asymmetry can be accounted for by the CME after
the electroweak phase transition~\cite{BoyFroRuc12} and for hypermagnetic
fields before this phase transition~\cite{GioSha98}.

The CME and CVE can be used in astrophysics mainly for the explanation
of the generation of strong magnetic fields with $B>10^{15}\,\text{G}$
observed in some compact stars, known as magnetars~\cite{KasBel17}.
The application of these chiral effects in the neutrino sector in
connection to the magnetars problem was made in Refs.~\cite{Yam16,Mas18}.
The explanation of pulsars kicks basing on the anomalous hydrodynamics
was proposed in Ref.~\cite{Kam16}. The study of the magnetic fields
generation in magnetars accounting for the CME in turbulent matter
was carried out in Ref.~\cite{SigLei16}.

We proposed various applications of the chiral phenomena for the
explanation of strong magnetic fields in magnetars in Refs.~\cite{DvoSem15,Dvo16a,DvoSem19,DvoSemSok20}.
Note that both the CME and the CSE were used in Refs.~\cite{DvoSem19,DvoSemSok20}
to tackle the problem of strong magnetic fields in magnetars. In the
present paper, we continue the series of our previous works for the
study of the magnetic fields evolution in compact stars. Now, we analyze
how the evolution of the magnetic helicity is driven by the electroweak
interaction of chiral fermions with dense inhomogeneous electroweak
matter inside a compact star.

In the present work, we study the plasma of ultrarelativistic electrons electrowealy interacting with nonrelativistic background matter composed of neutrons and protons. The electroweak interaction is accounted for in the Fermi approximation. The characteristics of this matter, such as the density and the macroscopic velocity can depend on coordinates. The electron gas is supposed to be degenerate, with the Fermi momentum $p_{\mathrm{F}e}$ being much greater than the electron mass $m_e$. Such a system may well exist in a neutron star (NS). The application of the chiral phenomena to electrons is justified since these particles are ultrarelativistic: $p_{\mathrm{F}e}\gg m_e$. Anomalous electric currents of protons are negligible since protons are nonrelativistic in NS. Hydrodynamic currents of neutrinos, which are ultrarelativistic particles, can receive an anomalous contribution. However, the fluxes of neutrinos are small for an old NS, studied in our work. Thus, we neglect the neutrino contribution.

This work is organized in the following way. First, in Sec.~\ref{sec:EVOL},
we formulate the dynamics of chiral fermions in dense inhomogeneous
matter, with the electroweak interaction being accounted for. The
evolution of chiral fermions was described approximately basing on
the Berry phase evolution. We find the effective actions and the kinetic
equations for chiral fermions in Sec.~\ref{sec:EVOL}. Then, in Sec.~\ref{sec:ELCURR},
we derive the corrections to the anomalous electric current and to
the Adler anomaly from the electroweak interaction with inhomogeneous
matter. Finally, we apply our results in Sec.~\ref{sec:APPL} for
the description of the magnetic helicity evolution in a rotating NS and compare our prediction for the period of the magnetic
cycle with the astronomical observations. We conclude in Sec.~\ref{sec:CONCL}.

\section{Approximate description of chiral plasma based on the Berry phase
evolution\label{sec:EVOL}}

We start with the consideration of a chiral fermion motion
in electroweak background matter. The interaction with the electromagnetic
field will be accounted for later. The Dirac equation for such a fermion, which is an electron,
reads,
\begin{equation}\label{eq:Direqcov}
  \mathrm{i}\dot{\psi}=
  \left[
    (\bm{\alpha}\hat{\mathbf{p}})+
    V_{\mathrm{R}}^{\mu}\gamma^{0}\gamma_{\mu}P_{\mathrm{R}}+
    V_{\mathrm{L}}^{\mu}\gamma^{0}\gamma_{\mu}P_{\mathrm{L}}
  \right]\psi,
\end{equation}
where $\hat{\mathbf{p}}=-\mathrm{i}\nabla$ is the momentum operator,
$\psi$ is the bispinor of the fermion, $\gamma^{\mu}=(\gamma^{0},\bm{\gamma})$,
$\bm{\alpha}=\gamma^{0}\bm{\gamma}$, and $\gamma^{5}=\mathrm{i}\gamma^{0}\gamma^{1}\gamma^{2}\gamma^{3}$
are the Dirac matrices, $P_{\mathrm{R,L}}=(1\pm\gamma^{5})/2$ are
the chiral projectors, and $V_{\mathrm{R},\mathrm{L}}^{\mu}=V_{\mathrm{R},\mathrm{L}}^{\mu}(\mathbf{x})=(V_{\mathrm{R},\mathrm{L}}^{0},\mathbf{V}_{\mathrm{R,L}})$
are the effective potentials for the interaction of the chiral projections
with inhomogeneous matter. The explicit expressions for $V_{\mathrm{R},\mathrm{L}}^{0}\sim G_{\mathrm{F}}n_{f}$
for electrons can be found in Ref.~\cite{DvoSem15}. Here $n_{f}$
is the density of background particles and $G_{\mathrm{F}}=1.17\times10^{-5}\,\text{GeV}^{-2}$
is the Fermi constant. If we consider arbitrarily moving unpolarized
background matter, then $\mathbf{V}_{\mathrm{R},\mathrm{L}}=V_{\mathrm{R},\mathrm{L}}^{0}\mathbf{v}$~\cite{DvoStu02},
where $\mathbf{v}(\mathbf{x})$ is the velocity of plasma, in which
all background particles are supposed to move as a whole. The density
of background matter can be also the function of coordinates, $n_{f}=n_{f}(\mathbf{x})$.
Note that $V_{\mathrm{R}}^{\mu}\neq V_{\mathrm{L}}^{\mu}$ since the
electroweak interaction violates parity.

Using the chiral representation for the Dirac matrices, we can rewrite
$\psi$ in the form of two chiral projections, $\psi^{\mathrm{T}}=(\psi_{\mathrm{R}},\psi_{\mathrm{L}})$.
Basing on Eq.~(\ref{eq:Direqcov}), one gets the wave equations for
$\psi_{\mathrm{R,L}}$,
\begin{equation}\label{eq:Direqchirproj}
  \mathrm{i}\dot{\psi}_{\mathrm{R,L}}=
  \left[
    \pm(\bm{\sigma}\hat{\mathbf{p}})+
    V_{\mathrm{R,L}}^{0}\mp(\bm{\sigma}\mathbf{V}_{\mathrm{R,L}})
  \right]
  \psi_{\mathrm{R,L}},
\end{equation}
where $\bm{\sigma}$ are the Pauli matrices.

The solution of Eq.~(\ref{eq:Direqchirproj}) for the arbitrary dependence
$V_{\mathrm{R},\mathrm{L}}^{\mu}(\mathbf{x})$ is barely possible.
However, we can analyze the evolution of $\psi_{\mathrm{R,L}}$ quasiclassically
using the concept of the Berry phase~\cite{Ber84}. For this purpose
we shift from the field theory to the classical mechanics, i.e. we take
that particles move along some trajectories. Thus we should appropriately
introduce the canonical coordinates $\{\mathbf{x}(t),\mathbf{p}(t)\}$.
In principle, the choice of the coordinates is voluntary. Therefore,
instead of the wave functions $\psi_{\mathrm{R,L}}(\mathbf{x},t)$,
depending on time $t$ and the coordinates $\mathbf{x}$, we consider
the dependence on $t$ only, $\psi_{\mathrm{R,L}}(t)$: $\psi_{\mathrm{R,L}}(\mathbf{x},t)\to\exp(\mathrm{i}\Phi)\psi_{\mathrm{R,L}}(t)$,
where $\Phi=\Phi(\mathbf{x},\mathbf{p})$ is the phase which depends
on the canonical coordinates. We do not fix the form of $\Phi$ yet.

Using Eq.~(\ref{eq:Direqchirproj}), we get that the time dependent
spinors $\psi_{\mathrm{R,L}}=\psi_{\mathrm{R,L}}(t)$ evolve as
\begin{equation}\label{eq:HRL}
  \mathrm{i}\dot{\psi}_{\mathrm{R,L}}=H_{\mathrm{R,L}}\psi_{\mathrm{R,L}},
  \quad
  H_{\mathrm{R,L}}=\pm(\bm{\sigma}\bm{\Pi})+V_{\mathrm{R,L}}^{0},
\end{equation}
where $\bm{\Pi}=\nabla\Phi-\mathbf{V}_{\mathrm{R,L}}$. The Hamiltonians
$H_{\mathrm{R,L}}$ in Eq.~(\ref{eq:HRL}) depend on $t$ only since
particles move along the trajectories $\mathbf{x}=\mathbf{x}(t)$.

Now we assume the adiabatic change of $H_{\mathrm{R,L}}$. In this
case, we take that~\cite{Vin90}
\begin{equation}\label{eq:psiTheta}
  \psi_{\mathrm{R,L}}(t)=\exp(-\mathrm{i}\Theta_{\mathrm{R,L}})u_{\mathrm{R,L}},
\end{equation}
where $\Theta_{\mathrm{R,L}}=\Theta_{\mathrm{R,L}}(t)$ is the Berry
phase~\cite{Ber84}, which is different for right and left particles,
and $u_{\mathrm{R,L}}$ are the constant spinors. We impose the following
constraints on $\psi_{\mathrm{R,L}}$~\cite{Vin90}:
\begin{equation}\label{eq:normcond}
  \psi_{\mathrm{R,L}}^{\dagger}\psi_{\mathrm{R,L}}=1,
  \quad
  \psi_{\mathrm{R,L}}^{\dagger}\dot{\psi}_{\mathrm{R,L}}=0,
\end{equation}
to fix their amplitudes and the phases.

To find $\Theta_{\mathrm{R,L}}$, first, we solve the Schr\"odinger
equation~\cite{Vin90},
\begin{equation}\label{eq:Schrequ}
  E_{\mathrm{R,L}}u_{\mathrm{R,L}}(\xi)=H_{\mathrm{R,L}}u_{\mathrm{R,L}}(\xi),
\end{equation}
in which the canonical variable $\xi_{a}=\{\mathbf{x},\mathbf{p}\}$,
$a=1,\dots,6$, is supposed to be constant: $\xi_{a}=\text{const}$.
The solution has the form,
\begin{equation}\label{eq:Energyu}
  E_{\mathrm{R,L}}=V_{\mathrm{R,L}}^{0}+|\bm{\Pi}|,
  \quad
  u_{\mathrm{R}}=
  \left(
    \begin{array}{c}
      e^{-\mathrm{i}\varphi/2}\cos\dfrac{\theta}{2} \\
      e^{\mathrm{i}\varphi/2}\sin\dfrac{\theta}{2}
    \end{array}
  \right),
  \quad
  u_{\mathrm{L}}=
  \left(
    \begin{array}{c}
      -e^{-\mathrm{i}\varphi/2}\sin\dfrac{\theta}{2} \\
      e^{\mathrm{i}\varphi/2}\cos\dfrac{\theta}{2}
    \end{array}
  \right),
\end{equation}
where the angles $\varphi$ and $\theta$ fix the direction of the vector $\bm{\Pi}$.

Using Eqs.~(\ref{eq:HRL})-(\ref{eq:Schrequ}), we get that the Berry
phases $\Theta_{\mathrm{R,L}}$ obey the equation~\cite{Vin90},
\begin{equation}\label{eq:EqTheta}
  \mathrm{i}\dot{\Theta}_{\mathrm{R,L}}=u_{\mathrm{R,L}}^{\dagger}
  \frac{\partial u_{\mathrm{R,L}}}{\partial\xi_{a}}\dot{\xi}_{a}.
\end{equation}
The solution of Eq.~(\ref{eq:EqTheta}) reads~\cite{Vin90}
\begin{equation}\label{eq:pathint}
  \Theta_{\mathrm{R,L}}(t)=\int_{C}A_{a}^{(\mathrm{R,L})}(\xi)\mathrm{d}\xi_{a},
\end{equation}
where $C$ is the particle trajectory in the phase space and
\begin{equation}\label{eq:Berryconngen}
  A_{a}^{(\mathrm{R,L})}=-\mathrm{i}u_{\mathrm{R,L}}^{\dagger}
  \frac{\partial u_{\mathrm{R,L}}}{\partial\xi_{a}},
\end{equation}
is the Berry connection~\cite{Vin90}.

It is convenient to represent the Berry connection in terms of two
three-dimensional vectors, $A_{a}^{(\mathrm{R,L})}=(\bm{\mathcal{B}}_{\mathrm{R,L}},\bm{\mathcal{A}}_{\mathrm{R,L}})$.
Then, using Eq.~(\ref{eq:Berryconngen}), we obtain that
\begin{align}\label{eq:ABRL}
  \mathcal{A}_{i}^{(\mathrm{R,L})} & =
  -\mathrm{i}u_{\mathrm{R,L}}^{\dagger}\frac{\partial u_{\mathrm{R,L}}}{\partial p_{i}}=
  -\mathrm{i}u_{\mathrm{R,L}}^{\dagger}\frac{\partial u_{\mathrm{R,L}}}{\partial\Pi_{j}}  
  \frac{\partial\Pi_{j}}{\partial p_{i}},
  \nonumber
  \\
  \mathcal{B}_{i}^{(\mathrm{R,L})} & =
  -\mathrm{i}u_{\mathrm{R,L}}^{\dagger}\frac{\partial u_{\mathrm{R,L}}}{\partial x_{i}}=
  -\mathrm{i}u_{\mathrm{R,L}}^{\dagger}\frac{\partial u_{\mathrm{R,L}}}{\partial\Pi_{j}}
  \frac{\partial\Pi_{j}}{\partial x_{i}}.
\end{align}
Therefore, instead of the integration along the particle trajectory,
we can rewrite Eq.~(\ref{eq:pathint}) in the form,
\begin{equation}\label{eq:ThetaAPi}
  \Theta_{\mathrm{R,L}}(t)=\int_{t_{0}}^{t}
  \left(
    \bm{\mathcal{B}}_{\mathrm{R,L}}\dot{\mathbf{x}}+
    \bm{\mathcal{A}}_{\mathrm{R,L}}\dot{\mathbf{p}}
  \right)
  \mathrm{d}t=
  \int_{t_{0}}^{t}\mathcal{A}_{\bm{\Pi}j}^{(\mathrm{R,L})}
  \left[
    \frac{\partial\Pi_{j}}{\partial p_{i}}\dot{p}_{i}+
    \frac{\partial\Pi_{j}}{\partial x_{i}}\dot{x}_{i}
  \right]
  \mathrm{d}t,
\end{equation}
where
\begin{equation}
  \mathcal{A}_{\bm{\Pi}j}^{(\mathrm{R,L})} =
  -\mathrm{i}u_{\mathrm{R,L}}^{\dagger}\frac{\partial u_{\mathrm{R,L}}}{\partial\Pi_{j}},
\end{equation}
which results from Eq.~(\ref{eq:ABRL}).

The wave function evolves as $\psi_{\mathrm{R,L}}(t)\sim\exp(-\mathrm{i}\Theta_{\mathrm{R,L}})$
in Eq.~(\ref{eq:psiTheta}). Recalling that the spinors $u_{\mathrm{R,L}}$
have the energies $E_{\mathrm{R,L}}$ in Eq.~(\ref{eq:Energyu})
and accounting for Eq.~(\ref{eq:ThetaAPi}), we can define the effective
energies,
\begin{equation}\label{eq:Eeff}
  E_{\mathrm{eff}}=E_{\mathrm{R,L}}+\mathcal{A}_{\bm{\Pi}j}^{(\mathrm{R,L})}  
  \left[
    \frac{\partial\Pi_{j}}{\partial p_{i}}\dot{p}_{i}+
    \frac{\partial\Pi_{j}}{\partial x_{i}}\dot{x}_{i}
  \right].
\end{equation}
Now we suppose that chiral fermions have the electric charge $e$
and interact with the external electromagnetic field $A^{\mu}=(A_{0},\mathbf{A})$.
Accounting for Eq.~(\ref{eq:Eeff}), we obtain the effective actions
for such particles,
\begin{align}\label{eq:SeffRL}
  S_{\mathrm{eff}}^{(\mathrm{R,L})}[\mathbf{x},\mathbf{p}]= &
  \int_{t_{0}}^{t}\mathrm{d}t
  \bigg\{
    \dot{\mathbf{x}}
    \left[
      \mathbf{p}+e\mathbf{A}(\mathbf{x})
    \right]-
    \mathcal{A}_{\bm{\Pi}j}^{(\mathrm{R,L})}
    \left[
      \frac{\partial\Pi_{j}}{\partial p_{i}}\dot{p}_{i}+
      \frac{\partial\Pi_{j}}{\partial x_{i}}\dot{x}_{i}
    \right]
    \notag
    \\
    & -
    \epsilon_{\mathrm{R,L}}(\mathbf{x})-eA_{0}(\mathbf{x})  
  \bigg\},
\end{align}
where, following Ref.~\cite{SonYam13}, we introduce the particle
energy dependence on coordinates $\epsilon_{\mathrm{R,L}}(\mathbf{x})$.

Up to now, the canonical coordinates and, hence $\bm{\Pi}$, are not
specified. We can choose $\{\mathbf{x},\mathbf{p}\}$ in such a way
that $\Phi=(\mathbf{px})$. In this case,
\begin{equation}
  \bm{\Pi}\equiv\mathbf{P}=\mathbf{p}-\mathbf{V}_{\mathrm{R,L}},
  \quad
  \frac{\partial\Pi_{j}}{\partial p_{i}}=\delta_{ij},
  \quad
  \frac{\partial\Pi_{j}}{\partial x_{i}}=-\frac{\partial V_{j}^{(\mathrm{R,L})}}{\partial x_{i}},
\end{equation}
and $\epsilon_{\mathbf{P}}^{(\mathrm{R,L})}(\mathbf{x})=V_{\mathrm{R,L}}^{0}+|\mathbf{P}|+\epsilon_{0}$,
where $\epsilon_{0}=\epsilon_{0}(\mathbf{x})$ arises owing to the
possible dependence of the electromagnetic field on coordinates.

The effective action in Eq.~(\ref{eq:SeffRL}) takes the form,
\begin{equation}\label{eq:SLRint}
  S_{\mathrm{eff}}^{(\mathrm{R,L})}=\int_{t_{0}}^{t}\mathrm{d}t
  \left\{
    \dot{\mathbf{x}}
    \left[
      \mathbf{p}+e\mathbf{A}(\mathbf{x})
    \right]-
    \bm{\mathcal{A}}_{\mathrm{R,L}}
    \left[
      \dot{\mathbf{p}}-(\dot{\mathbf{x}}\nabla)\mathbf{V}_{\mathrm{R,L}}
    \right]-
    \epsilon_{\mathbf{P}}^{(\mathrm{R,L})}(\mathbf{x})-eA_{0}(\mathbf{x})
  \right\},
\end{equation}
where, for brevity, we omit the index $\mathbf{P}$ in the Berry connection:
$\bm{\mathcal{A}}_{\mathbf{P}}^{(\mathrm{R,L})}\to\bm{\mathcal{A}}_{\mathrm{R,L}}$.
Using the fact that $\dot{\mathbf{P}}=\dot{\mathbf{p}}-(\dot{\mathbf{x}}\nabla)\mathbf{V}_{\mathrm{R,L}}$,
we can rewrite Eq.~(\ref{eq:SLRint}) as
\begin{equation}\label{eq:SRLfin}
  S_{\mathrm{eff}}^{(\mathrm{R,L})}[\mathbf{x},\mathbf{P}]=
  \int_{t_{0}}^{t}\mathrm{d}t
  \left\{
    \dot{\mathbf{x}}
    \left[
      \mathbf{P}+e\mathbf{A}_{\mathrm{eff}}(\mathbf{x})
    \right]-
    \bm{\mathcal{A}}_{\mathrm{R,L}}\dot{\mathbf{P}}-
    \epsilon_{\mathbf{P}}^{(\mathrm{R,L})}(\mathbf{x})-eA_{0}(\mathbf{x})
  \right\},
\end{equation}
where $\mathbf{A}_{\mathrm{eff}}=\mathbf{A}+\mathbf{V}_{\mathrm{R,L}}/e$.
Now we can use the canonical variables $\Xi_{a}=\{\mathbf{x},\mathbf{P}\}$.
This choice of canonical variables to construct a kinetic equation
of a fermion, electroweakly interacting with inhomogeneous background
matter, coincides with that in Ref.~\cite{Sil99}. One just needs
to replace $\mathbf{p}\leftrightarrow\mathbf{P}$ in Ref.~\cite{Sil99}.

%

Basing on Eq.~(\ref{eq:SRLfin}) and applying the approach developed
in Ref.~\cite{SonYam13}, we obtain the kinetic equation for the
distribution functions $f_{\mathrm{R,L}}(\mathbf{x},\mathbf{P},t)$
of right and left fermions in the form,
\begin{align}\label{eq:kineq}
  \frac{\partial f_{\mathrm{R,L}}}{\partial t}+ &
  \frac{1}{1+e\mathbf{B}_{\mathrm{eff}}\bm{\Omega}}
  \bigg\{
    \left[
      \mathbf{v}+e(\tilde{\mathbf{E}}\times\bm{\Omega}_{\mathbf{P}})+
      e(\mathbf{v}\cdot\bm{\Omega}_{\mathbf{P}})\mathbf{B}_{\mathrm{eff}}
    \right]
    \frac{\partial f_{\mathrm{R,L}}}{\partial\mathbf{x}}
    \nonumber
    \\
    & +
    \left[
      e\tilde{\mathbf{E}}+e(\mathbf{v}\times\mathbf{B}_{\mathrm{eff}})+
      e^{2}(\tilde{\mathbf{E}}\cdot\mathbf{B}_{\mathrm{eff}})\bm{\Omega}_{\mathbf{P}}
    \right]
    \frac{\partial f_{\mathrm{R,L}}}{\partial\mathbf{P}}
  \bigg\} =0,
\end{align}
where $\tilde{\mathbf{E}}=\mathbf{E}_{\mathrm{eff}}-e^{-1}\partial\epsilon_{\mathbf{P}}/\partial\mathbf{x}$,
$\mathbf{v}=\partial\epsilon_{\mathbf{P}}/\partial\mathbf{P}$, $\mathbf{E}_{\mathrm{eff}}=-\nabla A_{0}-\nabla V_{\mathrm{R,L}}^{0}/e$,
$\mathbf{B}_{\mathrm{eff}}=(\nabla\times\mathbf{A}_{\mathrm{eff}})$,
and $\bm{\Omega}_{\mathbf{P}}=(\nabla_{\mathbf{P}}\times\bm{\mathcal{A}}_{\mathrm{R,L}})$
is the Berry curvature. To derive Eq.~(\ref{eq:kineq}) we take $\epsilon_{\mathbf{P}}^{(\mathrm{R,L})}=V_{\mathrm{R,L}}^{0}+\epsilon_{\mathbf{P}}$,
where $\epsilon_{\mathbf{P}}(\mathbf{x})=|\mathbf{P}|+\epsilon_{0}(\mathbf{x})$.

Note that, if $\epsilon_{0}(\mathbf{x})=0$, Eq.~(\ref{eq:kineq})
coincides with that derived in Refs.~\cite{DvoSem17,DvoSem18a} for
a neutrino electroweakly interacting with electron-positron plasma
since $\tilde{\mathbf{E}}=\mathbf{E}_{\mathrm{eff}}=\mathbf{E}+\mathbf{E}_{\mathrm{EW}}$
and $\mathbf{B}_{\mathrm{eff}}=\mathbf{B}+\mathbf{B}_{\mathrm{EW}}$,
where $\mathbf{E}_{\mathrm{EW}}=-e^{-1}\nabla V_{\mathrm{R,L}}^{0}$
and $\mathbf{B}_{\mathrm{EW}}=e^{-1}(\nabla\times\mathbf{V}_{\mathrm{R,L}})$
are the effective corrections induced by the electroweak interaction to the
conventional electromagnetic field $F_{\mu\nu}=(\mathbf{E},\mathbf{B})$.

\section{Electric current and Adler anomaly in inhomogeneous matter accounting for the electroweak interaction\label{sec:ELCURR}}

In this section, we derive the corrections to the anomalous current
and to the Adler anomaly arising from the electroweak interaction
of chiral fermions with inhomogeneous background matter. We consider
the particular case of the rotating matter.

Defining the currents~\cite{SonYam13}
\begin{align}\label{eq:current}
  \mathbf{j}_{\mathrm{R,L}}= & -\int\frac{\mathrm{d}^{3}P}{(2\pi)^{3}}  
  \bigg[
    \epsilon_{\mathbf{P}}\frac{\partial f_{\mathrm{R,L}}}{\partial\mathbf{P}}+
    e
    \left(
      \bm{\Omega}_{\mathbf{P}}\cdot\frac{\partial f_{\mathrm{R,L}}}{\partial\mathbf{P}}
    \right)
    \epsilon_{\mathbf{P}}\mathbf{B}_{\mathrm{eff}}
    \notag
    \\
    & +
    \epsilon_{\mathbf{P}}
    \left(
      \bm{\Omega}_{\mathbf{P}}\times\frac{\partial f_{\mathrm{R,L}}}{\partial\mathbf{x}}  
    \right)+
    e
    \left(
      \mathbf{E}_{\mathrm{eff}}\times\bm{\Omega}_{\mathbf{P}}
    \right)f_{\mathrm{R,L}}
  \bigg],
\end{align}
and the number densities~\cite{SonYam13}
\begin{equation}\label{eq:dens}
  n_{\mathrm{R,L}}=\int\frac{\mathrm{d}^{3}P}{(2\pi)^{3}}
  \left[
    1+e
    \left(
      \mathbf{B}_{\mathrm{eff}}\cdot\bm{\Omega}_{\mathbf{P}}
    \right)
  \right]
  f_{\mathrm{R,L}}
\end{equation}
one gets that these quantities obey the equation~\cite{DvoSem17,DvoSem18a},
\begin{equation}\label{eq:conteq}
  \partial_{t}n_{\mathrm{R,L}}+(\nabla\cdot\mathbf{j}_{\mathrm{R,L}})=
  -e^{2}\int\frac{\mathrm{d}^{3}P}{(2\pi)^{3}}
  \left(
    \bm{\Omega}_{\mathbf{P}}\cdot\frac{\partial f_{\mathrm{R,L}}}{\partial\mathbf{P}}
  \right)
  \left(
    \mathbf{E}_{\mathrm{eff}}\cdot\mathbf{B}_{\mathrm{eff}}
  \right).
\end{equation}
The derivation of Eq.~(\ref{eq:normcond}) implies the validity of
the Maxwell equations for the effective electromagnetic fields~\cite{DvoSem17,DvoSem18a}:
$(\nabla\cdot\mathbf{B}_{\mathrm{eff}})=0$ and $\partial_{t}\mathbf{B}_{\mathrm{eff}}+(\nabla\times\mathbf{E}_{\mathrm{eff}})=0$.

Let us suppose that the distribution functions are homogeneous: $\partial f_{\mathrm{R,L}}/\partial\mathbf{x}=0$.
In this case, $\epsilon_{\mathbf{P}}=\mu_{\mathrm{R,L}}+\dotsb$,
where $\mu_{\mathrm{R,L}}$ are the chemical potentials of right and
left fermions and we keep only the zero order terms in the Fermi constant
$G_{\mathrm{F}}$. We obtain that the anomalous part of the electric
current $\mathbf{J}_{\mathrm{R,L}}=e\mathbf{j}_{\mathrm{R,L}}$ can
be expressed in the following way:
\begin{align}\label{eq:JRL}
  \mathbf{J}_{\mathrm{R,L}}= &
  -\mu_{\mathrm{R,L}}e^{2}\mathbf{B}_{\mathrm{eff}}
  \int\frac{\mathrm{d}^{3}P}{(2\pi)^{3}}
  \left(
    \bm{\Omega}_{\mathbf{P}}\cdot\frac{\partial f_{\mathrm{R,L}}}{\partial\mathbf{P}}
  \right) =
  \pm\frac{e^{2}}{4\pi^{2}}\mu_{\mathrm{R,L}}\mathbf{B}_{\mathrm{eff}}
  \notag
  \\
  & =
  \pm\frac{e^{2}}{4\pi^{2}}\mu_{\mathrm{R,L}}
  \left[
    \mathbf{B}+\frac{1}{e}(\nabla\times\mathbf{V}_{\mathrm{R,L}})
  \right].
\end{align}
The first term in Eq.~(\ref{eq:JRL}), $\mathbf{J}_{\mathrm{R,L}}\parallel\mu_{\mathrm{R,L}}\mathbf{B}$,
is the well known CME~\cite{Fuk08}. To derive Eq.~(\ref{eq:JRL})
we take the distribution function in the form,
\begin{equation}\label{eq:distrfun}
  f_{\mathrm{R,L}}(\mathbf{P})=
  \left\{
    \exp[\beta(|\mathbf{P}|-\mu_{\mathrm{R,L}})]+1
  \right\} ^{-1},
  \quad
  \frac{\partial f_{\mathrm{R,L}}}{\partial\mathbf{P}}=
  -\frac{\mathbf{P}}{|\mathbf{P}|}\frac{\beta\exp[\beta(|\mathbf{P}|-\mu_{\mathrm{R,L}})]}{
  \left\{
    \exp[\beta(|\mathbf{P}|-\mu_{\mathrm{R,L}})]+1
  \right\}^{2}
  }.
\end{equation}
where $\beta=1/T$ is the reciprocal temperature. In Eqs.~(\ref{eq:JRL})
and~(\ref{eq:distrfun}), we consider the degenerate plasma with
$\beta\to\infty$ and $\nabla_{\mathbf{P}}f_{\mathrm{R,L}}\to\delta(|\mathbf{P}|-\mu_{\mathrm{R,L}})$
and take into account that $\bm{\Omega}_{\mathbf{P}}=\pm\mathbf{P}/2|\mathbf{P}|^{3}$~\cite{DvoSem17}.

Analogously Eq.~(\ref{eq:JRL}) we can calculate the correction to
the Adler anomaly using Eq.~(\ref{eq:conteq}). First, we suppose
that the number density of background matter is homogeneous, i.e.
$\nabla V_{\mathrm{R,L}}^{0}=0$. In this case, $\mathbf{E}_{\mathrm{eff}}=\mathbf{E}$,
and
\begin{equation}
  \partial_{t}n_{\mathrm{R,L}}+(\nabla\cdot\mathbf{j}_{\mathrm{R,L}})=
  \pm\frac{e^{2}}{4\pi^{2}}
  \left(
    \mathbf{E}\cdot\mathbf{B}_{\mathrm{eff}}
  \right)=
  \pm\frac{e^{2}}{4\pi^{2}}
  \left[
    \left(
      \mathbf{E}\cdot\mathbf{B}
    \right)+
    \frac{1}{e}
    \left(
      \mathbf{E}\cdot\nabla\times\mathbf{V}_{\mathrm{R,L}}
    \right)
  \right].
\end{equation}
Integrating this expression over the volume $V$ and neglecting the
boundary effects, we obtain that
\begin{equation}\label{eq:n5gen}
  \frac{\mathrm{d}n_{5}}{\mathrm{d}t}=\frac{e^{2}}{2\pi^{2}V}\int
  \left[
    \left(
      \mathbf{E}\cdot\mathbf{B}
    \right)-
    \frac{1}{e}
    \left(
      \mathbf{E}\cdot\nabla\times\mathbf{V}_{5}
    \right)
  \right]
  \mathrm{d}^{3}x -
  \Gamma n_5,
\end{equation}
where $n_{5}=n_{\mathrm{R}}-n_{\mathrm{L}}$ is the chiral imbalance and
$\mathbf{V}_{5}=(\mathbf{V}_{\mathrm{L}}-\mathbf{V}_{\mathrm{R}})/2$. We have $(\nabla\times \mathbf{V}_{5}) = V_{5}\bm{\omega}$,
where $V_{5}=(V_{\mathrm{L}}^{0}-V_{\mathrm{R}}^{0})/2$ and $\bm{\omega}$
is the angular velocity.

In Eq.~\eqref{eq:n5gen}, we account for the fact that electrons are ultrarelativistic but not massless particles. Hence their helicity can be changed in collisions in plasma. Thus we introduce the term $-\Gamma n_5$ in the right hand side of Eq.~\eqref{eq:n5gen}. The spin flip rate $\Gamma\sim 10^{11}\,\text{s}^{-1}$ was computed for an old NS in Ref.~\cite{Dvo16b}.

The electroweak correction to the anomalous current $\delta\mathbf{J}_{\mathrm{anom}}=(\mathbf{J}_{\mathrm{R}}+\mathbf{J}_{\mathrm{L}})\parallel\nabla\times(\mathbf{V}_{\mathrm{R}}-\mathbf{V}_{\mathrm{L}})\sim V_{5}\bm{\omega}$
in Eq.~(\ref{eq:JRL}), in fact, reproduces the result of Ref.~\cite{Dvo15},
except the factor $1/2\pi$ which was missed in Ref.~\cite{Dvo15}
because of the incorrect normalization of the spinors there. The generation
of such a current was named the galvano-rotational effect in Ref.~\cite{Dvo15}.

\section{Astrophysical applications\label{sec:APPL}}

Let us consider the application of Eq.~(\ref{eq:n5gen}) in an astrophysical
plasma. We mentioned in Sec.~\ref{sec:ELCURR} that the spin flip rate is huge in NS. Hence, $\dot{n}_5\to0$ in Eq.~\eqref{eq:n5gen}. Using Eq.~\eqref{eq:n5gen}, we get that the chiral imbalance reaches the saturated value, 
\begin{equation}\label{eq:n5sat}
  n_5 \to n_5^{(\text{sat})}=\frac{e^{2}}{2\pi^{2}V\Gamma}\int
  \left[
    \left(
      \mathbf{E}\cdot\mathbf{B}
    \right)-
    \frac{1}{e}
    \left(
      \mathbf{E}\cdot\nabla\times\mathbf{V}_{5}
    \right)
  \right]
  \mathrm{d}^{3}x,
\end{equation}
We will show below that we can neglect $n_5^{(\text{sat})}$ in an old NS. Thus, the saturated densities of right and left electrons are equal.

If both $n_{5}$ and $\dot{n}_{5}$ are vanishing in Eq.~\eqref{eq:n5gen}, then, accounting for the fact that
\begin{equation}
  \frac{\mathrm{d}H}{\mathrm{d}t}=-2\int
  \left(
    \mathbf{E}\cdot\mathbf{B}
  \right)
  \mathrm{d}^{3}x,
\end{equation}
where
\begin{equation}\label{eq:maghel}
  H=\int
  \left(
    \mathbf{A}\cdot\mathbf{B}
  \right)
  \mathrm{d}^{3}x,
\end{equation}
is the magnetic helicity, we get the contribution of the electroweak interaction to the magnetic helicity evolution in the form,
\begin{equation}\label{eq:helevol}
  \left(
    \frac{\mathrm{d}H}{\mathrm{d}t}
  \right)_\mathrm{EW} =
  -2\frac{V_{5}}{e}\int
  \left(
    \mathbf{E}\cdot\bm{\omega}
  \right)
  \mathrm{d}^{3}x=-2\frac{V_{5}}{e\sigma}\int
  \left(
    \nabla\times\mathbf{B}\cdot\bm{\omega}
  \right)
  \mathrm{d}^{3}x.
\end{equation}
Here we take into account the Maxwell equations and the fact that
$\mathbf{E}=\mathbf{J}/\sigma=(\nabla\times\mathbf{B})/\sigma$, where
$\mathbf{J}$ is the ohmic current and $\sigma$ is the electric conductivity.

Equation~\eqref{eq:helevol} should be added to the contribution of the classical electrodynamics for the magnetic helicity evolution~\cite{Pri16},
\begin{equation}\label{eq:helevolclass}
  \left(
    \frac{\mathrm{d}H}{\mathrm{d}t}
  \right)_\mathrm{class} =
  -\frac{2}{\sigma}\int
  \left(
    \mathbf{J}\cdot\mathbf{B}
  \right)
  \mathrm{d}^{3}x+
  2\oint
  \left[
    ( \mathbf{B}\cdot\mathbf{A} ) ( \mathbf{v}\cdot\mathbf{n} ) -
    ( \mathbf{v}\cdot\mathbf{A} ) ( \mathbf{B}\cdot\mathbf{n} )
  \right]
  \mathrm{d}^{2}S,
\end{equation}
where $\mathbf{n}$ is the external normal to the stellar surface. The total rate of the helicity change is $\dot{H} = (\dot{H})_\mathrm{class} + (\dot{H})_\mathrm{EW}$, where the classical electrodynamics and the electroweak contributions are given in Eqs.~\eqref{eq:helevolclass} and~\eqref{eq:helevol} respectively. Since we concentrate mainly on the $(\dot{H})_\mathrm{EW}$ term in our work, we omit the subscript EW in the following.

Equation~(\ref{eq:helevol}) describes the new mechanism for the
helicity evolution owing to the electroweak interaction of chiral
fermions with inhomogeneous background matter. The inhomogeneity of
the interaction of chiral fermions is in the dependence of $\mathbf{V}_{\mathrm{R,L}}=V_{\mathrm{R,L}}^{0}\mathbf{v}$
on coordinates, which can take place if we consider rotating matter
with the velocity $\mathbf{v}(\mathbf{x})=\bm{\omega}\times\mathbf{x}$.
The anomalous contribution to the helicity evolution in Eq.~(\ref{eq:helevol})
occurs if $(\nabla\times\mathbf{B})\parallel\bm{\omega}$. Such a
situation can happen in a rotating star having a toroidal component
of the magnetic field. It should be noted that the influence of quantum
effects on the magnetic helicity evolution, analogous to that in Eq.~(\ref{eq:helevol}),
was studied in Ref.~\cite{DvoSem18b}.

\begin{figure}
  \centering
  \includegraphics[scale=0.5]{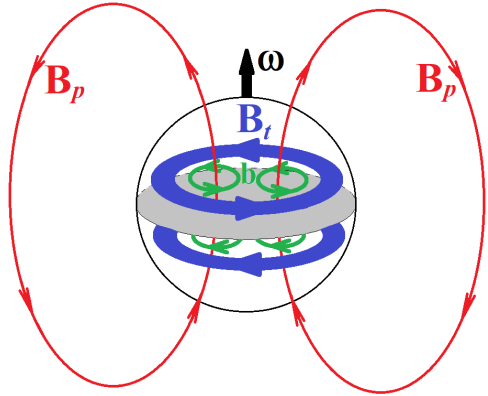}
  \caption{The schematic illustration of the stellar magnetic field configuration.
  The poloidal magnetic field $\mathbf{B}_{p}$, which can exist outside
  the star, is shown by red lines. The toroidal field $\mathbf{B}_{t}$
  is concentrated within two blue tori with opposite directions above
  and below the stellar equator, filled by the grey color. The fluctuations
  of the magnetic field in the form of green rings can exist in both
  the northern and southern hemispheres. The direction of the magnetic
  field $\mathbf{b}$ in these rings coincides with that for $\mathbf{B}_{t}$.\label{fig:starB}}
\end{figure}

We suppose that a compact star has the dipole configuration of the
magnetic field: the poloidal component $\mathbf{B}_{p}$ and two tori
of the toriodal field $\mathbf{B}_{t}$ with different directions in the 
opposite hemispheres, as schematically shown in Fig.~\ref{fig:starB}.
Note that a stellar magnetic field having only either a poloidal or
a toroidal component is unstable~\cite{BraNor06}.

Let us assume that the toroidal magnetic field in one of the hemispheres
of the star is a source of magnetic field vortices or rings, which
disattach from the torus. It can happen because of turbulence processes
taking place inside some compact rotating stars~\cite{And07}. Such
vortices are depicted in Fig.~\ref{fig:starB} in green. It should
be noted that the direction of the magnetic field in a vortex coincides
with that in a torus, in which $\mathbf{B}_{t}$ is concentrated.
Then, the electroweak mechanism, present in Eq.~\ref{eq:helevol},
acts on these vortices. The described process results in the change
of the stellar magnetic field. Namely, it leads to the helicity flux
through the stellar equator. Indeed, $\dot{H}$ has different signs
in opposite hemispheres since $\mathbf{B}_{t}$, as well as vortices,
has opposite directions in the northern and southern hemispheres.
It means that $\dot{H}_{\mathrm{tot}}=\dot{H}_{\mathrm{N}}+\dot{H}_{\mathrm{S}}=0$,
there $H_{\mathrm{N,S}}$ are the helicities of the northern and southern
hemispheres.

We suppose that the average strength of the field in a vortex is $\mathbf{b}$
and the average width of a flux tube of a vortex is $r$. Then, assuming
the hypothetical situation that all the magnetic field in the northern
hemisphere transforms into vortices and accounting for the magnetic
flux conservation, one gets that $br^{2}N=B_{t}R_{t}^{2}$, where
$N$ is the total number of vortices and $R_{t}$ is the radius of
a torus. Thus, we can estimate $(\nabla\times\mathbf{B})$ in Eq.~(\ref{eq:helevol})
as
\begin{equation}
  (\nabla\times\mathbf{B})\sim N\frac{b}{r}\sim\frac{B_{t}R_{t}^{2}}{r^{3}}.
\end{equation}
We consider, e.g., the northern hemisphere. Then, basing on Eq.~(\ref{eq:helevol}),
we have the helicity change rate in the form, 
\begin{equation}
  \dot{H}_{\mathrm{N}} \sim
  -\frac{4}{3}\pi B_{t}R_{t}^{2}\omega\frac{V_{5}}{e\sigma}\frac{R^{3}}{r^{3}},
\end{equation}
where $R\sim10\,\text{km}$ is the NS radius. The typical time of
the helicity change in this hemisphere is
\begin{equation}\label{eq:relaxtime}
  \tau\sim\frac{H_{\mathrm{N}}}{|\dot{H}_{\mathrm{N}}|}=
  \frac{3}{4\pi}\frac{e B_{p}\sigma r^{3}}{\omega V_{5}R},
\end{equation}
where we use the definition of the helicity alternative to that in
Eq.~(\ref{eq:maghel})~\cite{Ber99}, $H=2L\Phi_{1}\Phi_{2}$, where
$\Phi_{1,2}$ are the linked magnetic fluxes and $L=0,\pm1,\dotsc$
is the linkage number. Thus, in Eq.~(\ref{eq:relaxtime}), we take
that $H_{\mathrm{N}}\sim B_{t}R_{t}^{2}B_{p}R^{2}$ is the total helicity
in the northern hemisphere.

We consider the helicity change, proposed above, inside NS. In this
case, $V_{5}=G_{\mathrm{F}}n_{n}/2\sqrt{2}=6\,\text{eV}$~\cite{DvoSem15},
where $n_{n}=1.8\times10^{38}\,\text{cm}^{-3}$ is the neutron density
in NS. The application of the chiral phenomena for electrons in dense
matter of NS is justified since their Fermi momentum $p_{\mathrm{F}e}=(3\pi^{2}n_{e})^{1/3}\sim10^{2}\,\text{MeV}$ is much greater than the electron mass. Hence electrons are
ultrarelativistic. Here we take that the electron density $n_{e}$
in NS is about 5\% of $n_{n}$.

Then we take that $B_{p}=10^{12}\,\text{G}$ is the typical pulsar
magnetic field, $\omega=10^{3}\,\text{s}^{-1}$ is the angular velocity
of a millisecond pulsar, and $\sigma=10^{6}\,\text{GeV}$ is the electric
conductivity of the NS matter having the temperature $T\sim10^{7}\,\text{K}$~\cite{SchSht18}.
The magnetic flux tubes with $r\sim6\times10^{-5}\,\text{cm}$ were
predicted in Ref.~\cite{GusDom16} to exist in matter of some NSs.
Using the above values in Eq.~(\ref{eq:relaxtime}), we get that
$\tau=1.3\times10^{11}\,\text{s}=4\times10^{3}\,\text{yr}$.

In fact, we predict the magnetic helicity flux through the equator
driven by the electroweak interaction of chiral electrons with the
rotating matter of NS. The helicity flux through the stellar equator
was found in Ref.~\cite{Sor13} to be closely related to the cyclic
variation of the magnetic field in a star. For example, it is the
well known solar cycle with the period of $22\,\text{yr}$. Thus we
can conclude that $\tau\sim10^{3}\,\text{yr}$ is associated with
a periodic variation of the magnetic field in NSs. It should be noted
that such a cycle in NSs with $10^2\,\text{yr}<\tau<10^{4}\,\text{yr}$ was observed
in Ref.~\cite{Con07} by studying the spin-down of pulsars. The mechanism
proposed in our work is a possible explanation of the results of Ref.~\cite{Con07}.

At the end of this section, we demonstrate that we can neglect $n_5^{(\text{sat})}$ in Eq.~\eqref{eq:n5sat}. We mentioned above that the total helicity of NS is constant. The second term in the integrand in Eq.~\eqref{eq:n5sat}, $\sim \left( \mathbf{E}\cdot\nabla\times\mathbf{V}_{5}   \right)$, becomes important when turbulent rings of the magnetic field appear. Thus we can calculate the contribution to $n_5^{(\text{sat})}$ at the beginning of the helicity change process, i.e. from the first term in the integrand, $\sim \left( \mathbf{E}\cdot\mathbf{B} \right)$.
Taking into account that $\mathbf{E}=(\nabla\times\mathbf{B})/\sigma$, we obtain that $n_5^{(\text{sat})}\sim e^{2}B^2/2\pi^{2}R\sigma\Gamma$.
If we take the typical pulsar magnetic field $B = 10^{12}\,\text{G}$, the NS radius $R=10\,\text{km}$, the spin flip rate $\Gamma = 10^{11}\,\text{s}^{-1}$, and the conductivity $\sigma=10^{6}\,\text{GeV}$, we get that $n_5^{(\text{sat})}\sim 10^{12}\,\text{cm}^{-3}$.

The estimated value of $n_5^{(\text{sat})}$ should be compared with the seed chiral imbalance $n_5(0)$ which is created at the formation of NS. The seed chiral imbalance can appear in the parity violating direct Urca processes leading to the disappearance of left electrons from NS~\cite{DvoSem15}. The energy scale of this process is $\mu_5(0) = m_n - m_p \sim 1\,\text{MeV}$, where $\mu_5(0) = (\mu_\mathrm{R} - \mu_\mathrm{L})/2 = \pi^2 n_5(0)/p_{\mathrm{F}e}^2$ and $m_{n,p}$ are the masses of a neutron and a proton. Note that such a value of $\mu_5(0)$ was used in Refs.~\cite{DvoSem15,Dvo16a,DvoSemSok20}, where the problem of magnetars was tackled basing on the chiral phenomena. Taking that $p_{\mathrm{F}e}\sim10^{2}\,\text{MeV}$, we get that $n_5(0) \sim 10^{33}\,\text{cm}^{-3}$. We can see that $n_5^{(\text{sat})} \ll n_5(0)$ and, hence, we can neglect both $n_5$ and $\dot{n}_5$ in Eq.~\eqref{eq:n5gen} for an old NS.

Nevertheless there can be a back reaction of the evolving magnetic field on the chiral imbalance, which was recently studied in Ref.~\cite{DvoSemSok20}. Considering the full set of the anomalous hydrodynamics equations, we have found in Ref.~\cite{DvoSemSok20} that there is a spike in the $n_5$ evolution associated with the change of the polarity of a nascent NS. Concerning the problem considered in the present work, $n_5$ can grow sharply at the time $\tau$ estimated in Eq.~\eqref{eq:relaxtime}. However, this issue requires a separate study.

\section{Conclusion\label{sec:CONCL}}

In the present work, we have studied the chiral phenomena in inhomogeneous
plasma accounting for the electroweak interaction of chiral fermions
with background particles. In Sec.~\ref{sec:EVOL}, we have formulated
the wave equation for chiral fermions electroweakly interacting with
arbitrarily moving background matter with a nonuniform density. The
analysis of this equation has been performed basing on the Berry phase
approach. In Sec.~\ref{sec:EVOL}, we have derived the effective
actions and the kinetic equations for right and left particles which
generalize our results in Refs.~\cite{DvoSem17,DvoSem18a}.

Then, in Sec.~\ref{sec:ELCURR}, we have obtained the contribution
to the anomalous electric current and to the Adler anomaly from the
electroweak interaction of chiral fermions with inhomogeneous matter.
In this case, we studied a particular example of the rotating matter.
The contribution to the electric current is in agreement with the
result of Ref.~\cite{Dvo15}.

In Sec.~\ref{sec:APPL}, we have studied the application of our results
for the description of the magnetic fields evolution in dense matter
of a compact star. Supposing that the chiral imbalance is washed out
in interparticle collisions, we get the contribution to the evolution
of the magnetic helicity of a rotating NS owing to the electroweak
interaction of chiral electrons with background neutrons. This effect
results in the magnetic helicity flux through the stellar equator.
Assuming the formation of the magnetic vortices in the form of
rings because of the magnetic turbulence in NS, we have calculated the
typical time of the magnetic helicity change in one of the hemispheres
of NS. Note that the magnetic helicity of the whole star is constant.

The magnetic helicity flux through the equator of a star is believed
to be related to the cycle of the magnetic activity of such a star
analogous to the well known solar cycle. The typical time, calculated
in Sec.~\ref{sec:APPL}, is comparable with the period of the cyclic
electromagnetic activity of some pulsars reported in Ref.~\cite{Con07}.
Thus, the mechanism proposed in the present work is a possible explanation
of the observation in Ref.~\cite{Con07}.

\section*{Acknowledgments}

This work is supported by the Russian Science Foundation (Grant No.~19-12-00042).

\end{document}